\documentclass{article}

\usepackage{amsmath,amsfonts,amsthm}

\ifx\mathmr\undefined\let\mathmr=\mathrm\fi
\ifx\mathmb\undefined\let\mathmb=\mathbf\fi

\newcommand{\reals}{\mathbb{R}}
\newcommand{\F}{\mathcal{F}}
\newcommand{\U}{\mathcal{U}}
\renewcommand{\L}{\mathcal{L}}
\newcommand{\gl}{\mathfrak{gl}}
\newcommand{\id}{\mathmr{d}}
\renewcommand{\d}{\partial}
\newcommand{\dd}[1]{\frac{\d}{\d #1}}
\newcommand{\abs}[1]{\lvert#1\rvert}
\newcommand{\norm}[1]{\lVert{#1}\rVert}
\newcommand{\canform}{\boldsymbol{\theta}}
\newcommand{\connform}{\boldsymbol{\omega}}
\newcommand{\curvform}{\boldsymbol{\Omega}}
\newcommand{\connformLM}{\tilde{\boldsymbol{\omega}}}
\newcommand{\curvformLM}{\tilde{\boldsymbol{\Omega}}}

\renewcommand{\b}[1]{b\nobreakdash-\hspace{0pt}#1}
\newcommand{\pibar}{\bar{\pi}}
\newcommand{\LMbar}{\hspace{0.15em}\overline{\hspace{-0.15em}LM}}
\newcommand{\LpM}{L'\!M}
\newcommand{\GLp}{GL'}
\newcommand{\LpMbar}{\hspace{0.15em}\overline{\hspace{-0.15em}L'\!M}}
\newcommand{\dLM}{\partial LM}
\newcommand{\dLpM}{\partial L'\!M}
\newcommand{\Mbar}{\hspace{0.15em}\overline{\hspace{-0.15em}M}}
\newcommand{\dM}{\partial M}
\newcommand{\OMbar}{\hspace{0.18em}\overline{\hspace{-0.18em}OM}}
\newcommand{\dOM}{\partial OM}
\newcommand{\LplusM}{L^{\!+}\!M}
\newcommand{\LminusM}{L^{\!-}\!M}

\renewcommand{\sb}{\mskip-1.5\thinmuskip_}

\newtheorem{theorem}{Theorem}
\newtheorem{lemma}[theorem]{Lemma}
\newtheorem{proposition}[theorem]{Proposition}
\newtheorem{corollary}[theorem]{Corollary}

\begin{document}

\title{The Geometry of the Frame Bundle\\ over Spacetime} 
\author{Fredrik St{\aa}hl%
	\thanks{%
		\protect\parbox[t]{100mm}{%
			Department of Mathematics, University of Ume{\aa}, 
					S-901 87 Ume{\aa}, Sweden.\newline
			E-mail address: Fredrik.Stahl@math.umu.se}}}
\maketitle

\begin{abstract}
One of the known mathematical descriptions of singularities in General 
Relativity is the \b{boundary}, which is a way of attaching endpoints 
to inextendible endless curves in a spacetime.  The \b{boundary} of a 
manifold $M$ with connection $\Gamma$ is constructed by forming the 
Cauchy completion of the frame bundle $LM$ equipped with a certain 
Riemannian metric, the \b{metric} $G$.  We study the geometry of 
$(LM,G)$ as a Riemannian manifold in the case when $\Gamma$ is the 
Levi-Civit\`a connection of a Lorentzian metric $g$ on $M$.  In 
particular, we give expressions for the curvature and discuss the 
isometries and the geodesics of $(LM,G)$ in relation to the geometry 
of $(M,g)$.
\end{abstract}

\section{Introduction}
\label{sec:intro}

In general relativity, the concept of singularities is unavoidable.  
Known solutions to Einstein's field equations display a variety of 
non-trivial singularities, giving rise to such diverse phenomena as 
black holes, the big bang and topological anomalies.  The situation is 
very different from most other field theories, where singular 
solutions may be explained as artefacts of idealised modelling (like 
point charges) or differentiability restrictions (like shock waves), 
for example.  In order to study singularities, it is important to have 
a mathematical machinery that allows us to treat questions like 
convergence or divergence of physical quantities when approaching the 
singularity.  In other words, one would like to incorporate the 
singularities together with the regular spacetime points in some 
abstract set, equipped with a suitable topology that allows one to 
define statements such as `close to the singularity' in a 
mathematically precise sense.

One definition of this kind is the \b{boundary} 
\cite{Schmidt:b-boundary}.  Given a manifold $M$ with connection 
$\Gamma$, the \b{boundary} of $M$ is formed by constructing a suitable 
metric, the \b{metric} $G$, on the bundle of frames $LM$.  The pair 
$(LM,G)$ can then be viewed as a Riemannian manifold, and in 
particular, as a topological metric space.  The \b{boundary} is formed 
by taking the Cauchy completion of $(LM,G)$, and a projection then 
gives an extension $\Mbar$ of the original manifold $M$.  We leave the 
details to the next section.

The \b{boundary} construction has several drawbacks however, the most 
important being that the topology on the extended set $\Mbar$ is 
non-Hausdorff in general (see, e.g., \cite{Clarke:analysis-sing, 
Dodson:edge-geometry, Stahl:degeneracy}).  There has been some 
attempts to remedy the situation (see \cite{Dodson:new-completion, 
Dodson-Sulley:completion}), although they have not been entirely 
successful.

In order to obtain a more complete understanding of what goes wrong 
with the \b{boundary}, we need to understand the geometry of $(LM,G)$ 
better.  This is the subject at hand.  Since the object of interest is 
spacetime, we restrict attention to the case when $\Gamma$ is the 
Levi-Civit\`a connection of a Lorentzian metric $g$ on $M$.  

The outline of the paper is as follows: in \S\ref{sec:b-boundary}, we 
go through the essential steps of the \b{boundary} definition.  In 
\S\ref{sec:connection} and \S\ref{sec:curvature}, we calculate the 
connection and curvature of $(LM,G)$ and discuss some of the 
implications.  Finally, \S\ref{sec:isometries} and 
\S\ref{sec:geodesics} are devoted to a discussion of the isometries 
and the geodesics of $(LM,G)$ in relation to the corresponding 
structures on $(M,g)$.

\section{The b-boundary}
\label{sec:b-boundary}

We will attach an abstract boundary set to a Lorentzian manifold 
$(M,g)$, where $M$ is a smooth $n$-dimensional connected orientable 
Hausdorff manifold with a smooth metric $g$ of signature $n-2$. The 
case of interest in relativity theory is of course when $n=4$.

The construction of the \b{boundary} may be carried out in different 
bundles over $M$ (see~\cite{Schmidt:b-boundary}, 
\cite{Dodson:edge-geometry} or \cite{Hawking-Ellis} for some 
background).  It is often convenient to work with the bundle of 
pseudo-orthonormal frames $OM$, consisting of all pseudo-orthonormal 
frames at all points of $M$. In our case this leads to complications 
because of the amount of algebra involved, so we choose instead to 
construct the \b{boundary} via the bundle of general linear frames $LM$.

$LM$ is a principal fibre bundle with the general linear group $GL(n)$ 
on $\reals^n$ as its structure group.  We write the right action of an 
element $A\in GL(n)$ as $R_A:E\mapsto EA$ for $E\in LM$.  If $M$ is 
orientable, the frame bundle $LM$ has two connected components which 
we denote by $\LplusM$ and $\LminusM$, corresponding to frames with 
positive and negative orientation, respectively.  We will use the 
notation $\LpM$ for any one of these two components.  Clearly, any 
$A\in GL(n)$ which changes the orientation sets up a 1--1 
correspondence between $\LplusM$ and $\LminusM$, and we may regard the 
component of the identity $\GLp(n)\subset GL(n)$ as the structure 
group of $\LpM$.

The fibre bundle structure of $LM$ gives a canonical 1-form 
$\canform:T(LM)\to\reals^n$, and the connection 
corresponds to a connection form $\connform:T(LM)\to\gl(n)$, where 
$\gl(n)$ is the Lie algebra of $GL(n)$ \cite{Kobayashi-Nomizu-I}.  Let 
$\langle\cdot,\cdot\rangle_{\reals^n}$ and 
$\langle\cdot,\cdot\rangle_{\gl(n)}$ be Euclidian inner products with 
respect to fixed bases in $\reals^n$ and $\gl(n)$, respectively.  We 
define a Riemannian metric $G$ on $LM$, the \emph{\b{metric}} or 
\emph{Schmidt metric}, by
\begin{equation}\label{eq:Gdef}
	G(X,Y) := \langle\canform(X),\canform(Y)\rangle_{\reals^n}
	      	+ \langle\connform(X),\connform(Y)\rangle_{\gl(n)}.
\end{equation}
$G$ can be shown to be uniformly equivalent under a change of bases in 
$\reals^n$ and $\gl(n)$ 
\cite{Schmidt:b-boundary,Dodson:edge-geometry}.

If $\gamma$ is a curve in $LM$, the \emph{\b{length}} of $\gamma$ is 
the length of $\gamma$ with respect to the \b{metric} $G$, and is 
denoted by $l(\gamma)$.  Thus
\begin{equation}\label{def:b-length}
	l(\gamma) := \int	\Bigl( 
										\abs{\canform(\dot{\gamma})}^2 + 
										\norm{\connform(\dot{\gamma})}^2
										\Bigr)^{1/2} \,\id t,
\end{equation}
where $\abs\cdot$ and $\norm\cdot$ are the fixed Euclidian norms in 
$\reals^n$ and $\gl(n)$, respectively, and $\dot{\gamma}$ is the 
tangent of $\gamma$.

If $\gamma$ is horizontal, $\connform(\dot{\gamma})=0$ and $\gamma$ 
may be written as a pair $(\lambda,E)$ where $\lambda=\pi\circ\gamma$ 
is a curve in $M$ and $E$ is the parallel frame along $\lambda$ given 
by $\gamma$.  By definition, $\canform(\dot{\gamma})$ is the vector of 
the components of the tangent $\dot\lambda$ in the frame $E$.  So the 
\b{length} of a horizontal curve $\gamma$ is equivalent to the length 
measured in a parallel frame along $\pi\circ\gamma$ (which is called 
`generalised affine parameter length' in \cite{Hawking-Ellis}).  This 
motivates further study of the \b{metric}, since the presence of an 
endless curve with finite generalised affine parameter length is often 
taken as the criterion for a spacetime to be singular.

Following Schmidt \cite{Schmidt:b-boundary}, we now use the \b{metric} 
$G$ to construct a topological boundary of the base manifold $M$, 
providing endpoints for all endless curves with finite \b{length}.  
Since $(\LpM,G)$ is a Riemannian manifold, it is a metric space with 
topological metric $d$ (the \b{metric} distance function), and so the 
Cauchy completion $\LpMbar$ of $\LpM$ is well defined.  $\LpMbar$ is a 
complete metric space, and we define the boundary of $\LpM$ as 
$\dLpM:=\LpMbar\setminus \LpM$.

The topological metric $d$ then has a unique extension $\bar{d}$ to 
$\LpMbar$.  It can be shown that the action of $\GLp(n)$ is uniformly 
continuous on $(\LpM,G)$, viewed as a metric space
\cite{Schmidt:b-boundary,Dodson:edge-geometry}.  It follows that there 
is a unique uniformly continuous extension of the right action of 
$\GLp(n)$ to $(\LpMbar,G)$.

Justified by the above, we may now construct the topological space 
\begin{equation}
	\Mbar:=\LpMbar/\GLp(n),
\end{equation}
the set of orbits of $\GLp(n)$ in $\LpMbar$.  We can also define a 
continuous projection $\pibar:\LpMbar\to\Mbar$ taking a point in 
$\LpMbar$ to the corresponding orbit of $\GLp(n)$.  By definition, 
$\pibar$ coincides with $\pi$ on $\LpM$.  Thus $\pibar(\LpM)=M$ may be 
regarded as a subset of $\Mbar$, and we define the \emph{\b{boundary}} 
of $M$ as $\dM:=\Mbar\setminus M$.

It is important to emphasise that the topological structure of 
$\LpMbar$ may be quite complicated.  In particular, in many relevant 
cases $\LpMbar$ is non-Hausdorff 
\cite{Bosshard:b-boundary,Johnson:b-boundary,Clarke:analysis-sing, 
Stahl:degeneracy}.  This is related to the possibility of `fibre 
degeneracy', as the fibre bundle structure usually cannot be 
extended to $\LpMbar$.  A boundary orbit may even be a single point 
\cite{Clarke:sing-holonomy,Stahl:degeneracy}.

For the rest of this paper we will, somewhat sloppily, write $LM$ 
instead of $\LpM$ and $GL(n)$ instead of $\GLp(n)$.

\section{Cartan's equations and the connection}
\label{sec:connection}

It is clear from the definition \eqref{eq:Gdef} of $G$ that
\begin{equation}
	E^i:=\canform^i
	\qquad\text{together with}\qquad
	E^{\{i,i'\}}:=\connform^i\sb{i'}
\end{equation}
is a global orthonormal coframe field on $(LM,G)$.  If we let 
uppercase latin indices take values in the extended index set 
$\{i=1\dots n\}\cup\{\{i,i'\};\; i,i'=1\dots n\}$, then $E^I$ is a 
basis for the cotangent space $T^*(LM)$.  Note that the brackets 
denote ordered pairs and not unordered sets.  If we denote the 
connection form of $(LM,G)$ by $\connformLM$, applying Cartan's first 
equation with vanishing torsion gives
\begin{equation}\label{eq:Cartan1LM}
	\id E^I + \connformLM^I\sb{J} \wedge E^J = 0.
\end{equation}
We can also apply Cartan's equations to $\canform$ and $\connform$ 
on $M$, which gives
\begin{equation}\label{eq:Cartan1M}
	\id \canform^i + \connform^i\sb{j} \wedge \canform^j = 0
\end{equation}
and
\begin{equation}\label{eq:Cartan2M}
	\curvform^i\sb{j} = \id \connform^i\sb{j} 
	             + \connform^i\sb{k} \wedge \connform^k\sb{j},
\end{equation}
where $\curvform^i\sb{j} = \frac12 R^i\sb{jkl}\, \canform^k\canform^l$ 
is the curvature 2-form and $R^i\sb{jkl}$ are the frame components of 
the Riemann tensor of $(M,g)$.  Combining \eqref{eq:Cartan1LM} with 
\eqref{eq:Cartan1M} and \eqref{eq:Cartan2M} we get the system
\begin{equation}\label{eq:connLM}
\begin{split}
	\connformLM^i_J \wedge E^J &= \connform^i_j \wedge \canform^j, \\
	\connformLM^{\{i,i'\}}_J \wedge E^J &= 
		\connform^i_j \wedge \connform^j_{i'} - 
		\tfrac12\, R^i\sb{i'jk}\, \canform^j \canform^k.
\end{split}
\end{equation}
Since $G$ is a Riemannian metric, the metric condition is
\begin{equation}\label{eq:metriccond}
	\connformLM^I\sb{J} = - \connformLM^J\sb{I}.
\end{equation}
Solving (\ref{eq:connLM}--\ref{eq:metriccond}) for $\connformLM$ we get
\begin{equation}\label{eq:connLMsoln}
\begin{split}
	\connformLM^i\sb{j} &= 
			\tfrac12 ( \connform^i\sb{j} - \connform^j\sb{i} )
	  - \tfrac12 \sum_{k,l} R^k\sb{lij} \connform^k\sb{l}, \\
	\connformLM^i\sb{\{j,j'\}} &= -\connformLM^{\{j,j'\}}\sb{i} = 
		- \tfrac12( \delta^i\sb{j} \canform^{j'} + \delta^i\sb{j'} \canform^j )
		- \tfrac12 \sum_{k} R^j\sb{j'ik} \canform^k, \\
	\connformLM^{\{i,i'\}}\sb{\{j,j'\}} &= 
		\tfrac12\Bigl[
				\delta^i\sb{j} ( \connform^{i'}\sb{j'} - \connform^{j'}\sb{i'} )
			+ \delta^{i'}\sb{j'} ( \connform^i\sb{j} - \connform^j\sb{i} )
			+ \delta^{i'}\sb{j} \connform^i\sb{j'}
			- \delta^i\sb{j'} \connform^j\sb{i'}
		\Bigr].
\end{split}
\end{equation}
We have left the index positions fixed and written out the summations 
explicitly to avoid confusing the two metrics involved.

\section{The curvature}
\label{sec:curvature}

Our objective is now to compute the curvature of $(LM,G)$ expressed 
in the basis $E^I$.  If we denote the curvature form on $LM$ by 
$\curvformLM$, Cartan's second equation on $LM$ gives
\begin{equation}\label{eq:Cartan2LM}
	\curvformLM^I\sb{J} = \id\connformLM^I\sb{J} 
									+ \connformLM^I\sb{K} \wedge \connformLM^K\sb{J}.
\end{equation}
In order to calculate $\id\connformLM$ we first need an expression for 
$\id R^i\sb{jkl}$, which is given by the following lemma.  The result 
is not new of course (see, e.g., \cite{Bradley-Marklund:invariants}), 
but we still provide a proof for completeness.

\begin{lemma}\label{la:dR}
The exterior derivatives of the frame components of the Riemann 
tensor, viewed as functions on $LM$, are
\begin{equation}
	\id R^i\sb{jkl} = 
		- \,R^m\sb{jkl} \connform^i\sb{m}
		+ R^i\sb{mkl} \connform^m\sb{j}
		+ R^i\sb{jml} \connform^m\sb{k}
		+ R^i\sb{jkm} \connform^m\sb{l}
		+ R^i\sb{jkl;m} \canform^m.
\end{equation}
Here $R^i\sb{jkl;m}$ denote the frame components of the covariant 
derivative of the Riemann tensor of $(M,g)$.
\end{lemma}

\begin{proof}
Let $E_i$ be the standard horizontal vector fields dual to 
$\canform^i$, fixed by the choice of basis for $\reals^n$.  Similarly, 
let $E_{\{i,i'\}}$ be the fundamental vertical vector fields dual to 
$\connform^i_{i'}$, corresponding to the choice of basis for 
$\gl(n)$. Clearly $E_I=\{E_i,E_{\{i,i'\}}\}$ is a basis for $T(LM)$. 
Moreover, it is the unique dual of $E^I$. From the definition of the 
exterior derivative,
\begin{equation}\label{eq:dRdef}
	\id R^i\sb{jkl} =
		\sum_m E_m(R^i\sb{jkl})\,\canform^m 
	+ \sum_{m,m'} E_{\{m,m'\}}(R^i\sb{jkl})\,\connform^m\sb{m'}.
\end{equation}

To proceed further we introduce coordinates on $LM$ as follows.  Let 
$p\in M$ and let $x^i$ be coordinates for $M$ on a 
neighbourhood $\U$ of $p$.  Given a frame $F=(F_i)$ at a 
point $q\in\U$, we may express each $F_i$ as
\begin{equation}
	F_i = \Bigl(\dd{x^a}\Bigr) X^a\sb{i}.
\end{equation}
The determinant of the matrix $X$ is nonzero, so we may use 
$(x^a,X^b\sb{i})$ as coordinates for $LM$ on $\pi^{-1}(\U)$.

The coordinate expressions for the horizontal vector fields $E_i$ and 
the vertical vector fields $E_{\{i,i'\}}$ are then 
\begin{align}
	\label{eq:horvf}
	E_i &= \Bigl( \dd{x^a} 
								- \Gamma^b\sb{ac} X^c\sb{j} \dd{X^b\sb{j}} \Bigr) X^a\sb{i}, \\
	\label{eq:vervf}
	E_{\{i,i'\}} &= \Bigl(\dd{\displaystyle X^a\sb{i'}}\Bigr) X^a\sb{i},
\end{align}
where $\Gamma^b\sb{ac}$ are the Christoffel symbols of $(M,g)$ in the 
coordinates $x^a$ \cite{Kobayashi-Nomizu-I}.  Now the Riemann tensor 
frame components $R^i\sb{jkl}$ are related to the coordinate 
components $R^a\sb{bcd}$ by
\begin{equation}\label{eq:Rtransform}
	R^i\sb{jkl} = (X^{-1})^i\sb{a}R^a\sb{bcd}X^b\sb{j} X^c\sb{k} X^d\sb{l},
\end{equation}
where $X^{-1}$ is the inverse of $X$.  Applying 
(\ref{eq:horvf}--\ref{eq:vervf}) to \eqref{eq:Rtransform} and 
inserting into \eqref{eq:dRdef} then gives the desired result.
\end{proof}

With the help of \eqref{eq:connLMsoln}, 
(\ref{eq:Cartan1M}--\ref{eq:Cartan2M}) and Lemma~\ref{la:dR}, it is 
possible to solve \eqref{eq:Cartan2LM} for the curvature form 
$\curvformLM$, and then calculate the Riemann tensor, the Ricci tensor 
and the curvature scalar.  It is a trivial but long and tedious 
exercise, so we will not describe it here.  Some of the results are 
given in Appendix~\ref{app:curvature}, here we just give the 
expression for the curvature scalar $\tilde{R}$:
\begin{equation}\label{eq:curvscalarLM}
	\tilde{R} = -\,\tfrac12 n^2 (n+3) 
							- \tfrac14 \sum_{i,j,k,l} (R^i\sb{jkl})^2
							+ \sum_{i} R_{ii},
\end{equation}
where $R^i\sb{jkl}$ and $R_{ij}$ are the frame components of the 
Riemann and Ricci tensor of $(M,g)$, respectively.

What is the relevance of \eqref{eq:curvscalarLM} in relation to 
singularities?  A \b{incomplete} endless curve $\lambda$ in $M$ has an 
endpoint $p$ on the \b{boundary} $\dM$.  The horizontal lift $\gamma$ of 
$\lambda$ has finite \b{length} and ends at $\dLM$.  In many cases, some 
frame component of the Riemann tensor will diverge along $\gamma$ 
(c.f. \cite{Hawking-Ellis,Clarke:analysis-sing, 
Clarke-Schmidt:state-of-the-art,Ellis-Schmidt:singular}).  Using the 
terminology of \cite{Ellis-Schmidt:singular}, we say that $p$ is a 
\emph{curvature singularity}.  We treat the case when the divergence 
is unbounded.

\begin{proposition}\label{pr:scalardiv}
Let $\gamma$ be a horizontal curve ending at $\dLM$, and suppose that 
some frame component of the Riemann tensor of $(M,g)$ tends to 
$\pm\infty$ along $\gamma$.  Then the curvature scalar $\tilde{R}$ of 
$(LM,G)$ tends to $-\infty$ along $\gamma$.
\end{proposition}

\begin{proof}
The last term in \eqref{eq:curvscalarLM} may be written as
\begin{equation}
	\sum_{i,j} R^i\sb{jij},
\end{equation}
which is clearly dominated by the second term if some 
$R^i\sb{jkl}\to\pm\infty$.
\end{proof}

We now turn to the case when the frame components of the Riemann 
tensor are bounded on a curve ending at the boundary point.  In this 
case the obstruction to extending $M$ as a spacetime could be either 
some oscillatory divergence of the curvature, or purely topological.  
One might ask if it is possible that $(LM,G)$ can be extended as a 
Riemannian manifold even if $(M,g)$ is inextendible.  We answer this 
question in the case when the boundary fibre is totally degenerate, in 
a sense which we now specify.

The \b{boundary} construction outlined in \S\ref{sec:b-boundary} can 
also be carried out using the bundle of pseudo-orthonormal frames 
$OM$.  $OM$ is a fibre bundle with the Lorentz group $\L$ as its 
structure group, and there is a natural inclusion $OM\subset LM$ such 
that $OM$ is a reduced subbundle of $LM$ \cite{Kobayashi-Nomizu-I}.  
The \b{metric} is defined on $OM$ by \eqref{eq:Gdef}, i.e., exactly in 
the same way as for $LM$.  Since the connection in $OM$ is simply the 
reduction of the connection in $LM$, $(OM,G)$ is a Riemannian 
submanifold of $(LM,G)$.  We write $\OMbar$ and $\dOM$ for the Cauchy 
completion and boundary of $OM$, respectively.  Then an alternative 
definition (see \cite{Dodson:edge-geometry,Friedrich:b-boundary}) of 
the \b{boundary} is $\dM:=\Mbar\setminus M$ with
\begin{equation}
	\Mbar:=\OMbar/\L.
\end{equation}

In \cite{Stahl:degeneracy}, it was shown that for many exact solutions 
in general relativity, the `fibre' (or, more correctly, the orbit of 
the extended group action) over a point $p\in\dM$ in $OM$ is totally 
degenerate, i.e.,\ a single point.  Since $\dOM\subset\dLM$, the fibre 
in $LM$ is degenerate as well, though possibly not completely.

\begin{proposition}\label{pr:extension}
Suppose that $(LM,G)$ is (locally) extendible through 
$\pibar^{-1}(p)$, where $p\in\d M$, and that the corresponding boundary 
fibre in $\OMbar$ is completely degenerate.  Then $(M,g)$ is 
asymptotically a conformally flat Einstein space, i.e.\ the curvature 
$R$ of $(M,g)$ is given by $R^i\sb{jkl}=\tfrac16 R g^i\sb{[k}g_{l]\,j}$ 
in the limit at $p$.
\end{proposition}

\begin{proof}
Let $\lambda\colon[0,1]\to\OMbar\subset\LMbar$ be a horizontal curve 
ending at $\pibar^{-1}(p)$ such that the restriction to $[0,1)$ is 
contained in $OM$.  Since $(LM,G)$ is extendible through 
$\pibar^{-1}(p)$, the curvature scalar $\tilde{R}$ must have a well 
defined limit along $\lambda(t)$ as $t\to1$.  The restriction of 
$\pibar^{-1}(p)$ to $\OMbar$ is a single point, so the limit of 
$\tilde{R}$ must be invariant under the action of any Lorentz 
transformation $L\in\L$, changing the curve according to 
$\lambda\mapsto\lambda L$.

Now the curvature scalar $\tilde{R}$ is given by 
\eqref{eq:curvscalarLM}, which may be rewritten as
\begin{equation}\label{eq:curvscalarLM2}
	\tilde{R} = -\,\tfrac12 n^2 (n+3) - \tfrac14 I + R
							- 2 \sum_{\alpha,\beta,\gamma} (R^1\sb{\alpha\beta\gamma})^2
							+ 2 R_{11},
\end{equation}
where $1$ is the timelike index, greek indices go from 2 to $n$, $R$ 
is the curvature scalar and $I$ is the scalar invariant 
$R^{ijkl}R_{ijkl}$ of $(M,g)$.

Since $I$ and $R$ are scalar invariants, we only need to consider the 
last two terms in \eqref{eq:curvscalarLM2}.  Given the frame $E$ at 
$\lambda(t)$, we apply Lorentz transformations in the $n-1$ timelike 
planes spanned by $E_1$ and $E_\alpha$, $\alpha=2,3,\dots,n$.  After 
some algebra we find that \eqref{eq:curvscalarLM2} is invariant if and 
only if
\begin{align}
	\label{eq:Rinvar1}
	R^1\sb{\alpha\beta\gamma} &= R^\beta\sb{\alpha1\gamma}, \\
	\label{eq:Rinvar2}
	R^1\sb{\alpha1\beta} &= R^\gamma\sb{\alpha\gamma\beta}, \\
	\label{eq:Rinvar3}
	R^1\sb{\alpha1\beta} &= 0 \qquad\quad\text{if $\alpha\ne\beta$}, \\
	\label{eq:Rinvar4}
	R^\alpha\sb{\beta\gamma\epsilon} &= 0 
		\qquad\quad\text{if $\alpha\notin\{\beta,\gamma,\epsilon\}$}.
\end{align}
From \eqref{eq:Rinvar2} and \eqref{eq:Rinvar3}, the Ricci tensor is 
given by 
\begin{equation}
	R_{ij} = \tfrac14 R g_{ij},
\end{equation}
which is the condition for an Einstein space \cite{Spivak:IV}.
Applying the first Bianchi identity to \eqref{eq:Rinvar1} and 
permuting the indices shows that $R^1\sb{\alpha\beta\gamma}=0$. Thus 
(\ref{eq:Rinvar1}--\ref{eq:Rinvar4}) implies
\begin{equation}
	R^i\sb{jkl}= \tfrac16 R g^i\sb{[k}g_{l]\,j},
\end{equation}
which means that the Weyl tensor vanishes.
\end{proof}

\section{Isometries}
\label{sec:isometries}

\subsection{Horizontal isometries}
\label{sec:horiso}

We seek isometries of $(LM,G)$ which have horizontal orbits.  Any 
transformation $\varphi$ of $M$ induces an automorphism 
$\tilde\varphi$ of $LM$ taking a frame
$E=(E_1,E_2,\dots,E_n)$ at $p\in M$ to the frame 
$\tilde\varphi=(\varphi_*E_1,\varphi_*E_2,\dots,\varphi_*E_n)$
at $\varphi(p)\in M$.  A transformation of $(M,g)$ is said to be 
\emph{affine} if it preserves the connection.  Obviously, the group of 
affine transformations includes the isometry group.  From 
\cite{Kobayashi-Nomizu-I} we have the following.
\begin{proposition}\label{pr:affinexform}\ 
\par\nopagebreak\noindent
\textnormal{(1)} For any $\varphi$, $\tilde\varphi$ leave the 
canonical 1-form $\canform$ invariant.  Moreover, any fibre-preserving 
automorphism of $LM$ that leaves $\canform$ invariant is induced by a 
transformation of $M$.
\par\noindent
\textnormal{(2)} The fibre-preserving automorphisms of $LM$ that 
leaves both the canonical 1-form $\canform$ and the connection form 
$\connform$ invariant are exactly those that are induced by the affine 
transformations of $(M,g)$.
\end{proposition}
It is worth noting that $\tilde\varphi$ maps $OM$ into itself if and 
only if $\varphi$ is an isometry of $(M,g)$ \cite{Kobayashi-Nomizu-I}.  
From Proposition~\ref{pr:affinexform} we draw the following conclusion 
which is apparent from the definition \eqref{eq:Gdef} of the \b{metric}.
\begin{corollary}\label{co:horiso}\ \par\nopagebreak\noindent
Any affine transformation $\varphi$ of $(M,g)$ induces an isometry 
$\tilde\varphi$ of $(LM,G)$.
\end{corollary}

There might of course be other isometries of $(LM,G)$ induced by 
non-affine transformations of $(M,g)$.  They are characterised by the 
property that they preserve the inner product 
$\langle\connform,\connform\rangle_{\gl(n)}$.  This includes 
transformations that change the connection form according to 
$\connform\mapsto A\,\connform B$ where $A$ and $B$ belong to the 
rotation group $O(n)$.

\subsection{Vertical isometries}
\label{sec:veriso}

Our objective is now to find all vertical isometries of $(LM,G)$, 
i.e., the isometries where each orbit is contained in a single fibre 
of $LM$.  Assume that $V=V^I E_I$ is a vertical Killing vector field.  
Then the corresponding covector $V_I$ is given by
\begin{equation}
	V_i=0 \qquad\text{and}\qquad V_{\{i,i'\}}=a_{ii'},
\end{equation}
for some function $a$ from $LM$ to the Lie algebra $\gl(n)$, and 
satisfies the Killing equation
\begin{equation}\label{eq:verKilling}
	V_{(I;J)}=0.
\end{equation}
Here the semicolon denotes a covariant derivative with respect to $G$ 
and the parentheses denote symmetrisation.

From the expressions \eqref{eq:connLMsoln} for the connection form 
$\connformLM$ we may identify the connection coefficients in the frame 
$E$ on $LM$.  Then \eqref{eq:verKilling} becomes
\begin{gather}
	\label{eq:verKilling1}
	a_{(ij)} = 0, \\
	\label{eq:verKilling2}
	E_i (a_{jk}) = 0, \\
	\label{eq:verKilling3}
	E_{\{i,j\}}(a_{kl}) + E_{\{k,l\}}(a_{ij}) 
	- a_{(ik)}\delta_{jl} + a_{(jl)}\delta_{ik} = 0.
\end{gather}
From \eqref{eq:verKilling1}, $a$ is skew, and from 
\eqref{eq:verKilling2}, $a$ is a function of the fibre coordinates 
only. Loosely speaking, $a$ is the same on all fibres (locally). 
Thus \eqref{eq:verKilling3} gives
\begin{equation}\label{eq:verKilling3new}
	E_{\{i,j\}}(a_{kl}) + E_{\{k,l\}}(a_{ij}) = 0.
\end{equation}
This equation is obviously fulfilled if $a$ is a constant element in 
$\mathfrak{o}(n)$, the Lie algebra of the rotation group $O(n)$.  Then 
$V$ is the Killing vector field of a global isometry generated by the 
right action $R_A$ of $A=\exp{a}\in O(n)$.  (This can of course be 
seen directly from the definition \eqref{eq:Gdef} of the \b{metric}, 
using the transformation properties of $\canform$ and $\connform$ 
under the action $R_A$.)

Instead of searching for non-constant solutions to 
\eqref{eq:verKilling3new}, we reformulate the problem as finding local 
isometries of a single fibre $\F$, which is warranted by 
\eqref{eq:verKilling2}.  Since $\F$ is isomorphic to $GL(n)$, we may 
view $a_{ij}$ as the Killing vector field of a local isometry of 
$GL(n)$ with the metric $G_{\F}$ induced by $G$.  By the definition of 
the connection form $\connform$, $G_{\F}$ is invariant under the 
\emph{left} action of $GL(n)$, but not necessarily under the right 
action.  To study this in more detail, we introduce coordinates on 
$\F$ as in \S\ref{sec:curvature}.  Let $x^i$ be coordinates in a 
neighbourhood of the base point $\pi(\F)\in M$.  Then any frame 
$F=(F_i)\in\F$ can be expressed by $F_i = (\dd{x^a})X^a\sb{i}$ for some 
$X\in GL(n)$, and the map $\varphi\colon\F\to GL(n)$ given by 
$F\mapsto X$ is an isomorphism.  The restriction of the connection 
form $\connform$ to the fibre is
\begin{equation}\label{eq:connformF}
	\connform^i\sb{j} = (X^{-1})^i\sb{k}\, \id X^k\sb{j},
\end{equation}
and $G_{\F}$ is given by 
\begin{equation}\label{eq:GF}
	G_{\F}(X,Y)=\langle\connform(X),\connform(Y)\rangle_{\gl(n)}. 
\end{equation}
Apparently, we may view $G_{\F}$ as a metric on $GL(n)$.  The 
invariance of $G_{\F}$ under left translations $X\to AX$ is apparent 
from \eqref{eq:connformF}.  But we also see that $G_{\F}$ is not 
invariant under the right action of $GL(n)$.

Under $\varphi$, the orthonormal basis $E_{\{i,j\}}$ of\, $T(\F)$ is 
mapped to the orthonormal basis $e_{ij}$ of $\gl(n)$ used when defining 
the inner product $\langle\cdot,\cdot\rangle_{\gl(n)}$.  From the fact 
that $a$ is skew and \eqref{eq:verKilling3new} we get
\begin{equation}
	(e_{ij} + e_{ji})(a_{kl}) = 0,
\end{equation}
which means that $a$ is completely specified by its value on 
$O(n)\subset GL(n)$.  Also, from \eqref{eq:connformF} and 
\eqref{eq:GF}, the restriction of $G_{\F}$ to $O(n)$ is invariant 
under both left and right translations of $O(n)$.  Since $O(n)$ is a 
compact Lie group, there is essentially only one choice of a 
bi-invariant metric on $O(n)$ \cite{Gallot-Hulin-Lafontaine}.  We have 
thus arrived at a complete characterisation of the vertical isometries 
of $(LM,G)$.
\begin{proposition}\label{pr:veriso}
The group of (local) vertical isometries of $(LM,G)$ is isomorphic to 
the group of (local) isometries of $O(n)$ equipped with the 
`canonical' bi-invariant metric.
\end{proposition}

\section{Geodesics}
\label{sec:geodesics}

Suppose that $\gamma$ is a geodesic of $(LM,G)$, with tangent vector 
$v^i E_i + V^{ij} E_{\{i,j\}}$ and affine parameter $t$.  Reading off 
the connection coefficients from \eqref{eq:connLMsoln} and inserting 
into the geodesic equation gives
\begin{align}
	\label{eq:LMgeo1}
	\dot{v}^i &= \sum_j V^{ji}v^j + \sum_{j,k,l} R^k\sb{lij}V^{kl}v^j, \\
	\label{eq:LMgeo2}
	\dot{V}^{ii'} &= - v^i v^{i'} 
									+ \sum_{k}\bigl( V^{ki}V^{ki'} - V^{ik}V^{i'k} \bigr),
\end{align}
where the dot denotes an ordinary derivative with respect to $t$.  
Note that the right hand side of \eqref{eq:LMgeo2} is symmetric, which 
means that the skew part of $V^{ii'}$ must be constant.

\subsection{Vertical geodesics}
\label{sec:vergeo}

As we saw in \S\ref{sec:veriso}, any fibre $\F$ with the metric $G_{\F}$ 
induced by $G$ is isometric to $GL(n)$ with the metric generated by 
the inner product $\langle\cdot,\cdot\rangle_{\gl(n)}$. Thus the 
vertical geodesics are simply the geodesics of $GL(n)$ with respect to 
this metric, so the vertical geodesics are not coupled to the 
geometry of $(M,g)$ at all. Still, we clarify a few points.

If $v^{i}\equiv0$, \eqref{eq:LMgeo1} is automatically satisfied and 
\eqref{eq:LMgeo2} becomes
\begin{equation}\label{eq:vergeo}
	\dot{V}^{ii'} = \sum_{k}\bigl( V^{ki}V^{ki'} - V^{ik}V^{i'k} \bigr).
\end{equation}
We see that if $V$ is constant and skew, it satisfies 
\eqref{eq:vergeo}.  In other words, orbits of $R_{A}$ are geodesics if 
and only if $A\in O(n)$.  This is not surprising since $O(n)$ is a 
compact Lie group, and $G_{\F}$ corresponds to the canonical 
bi-invariant metric on $O(n)$.  Thus $(O(n),G_{\F})$ is a normal 
homogeneous space, and it is a property of such spaces that the 
geodesics are given by orbits of right actions 
\cite{Gallot-Hulin-Lafontaine}.  Note that this means that 
$(O(n),G_{\F})$ is a totally geodesic submanifold of $(GL(n),G_{\F})$.

Constant symmetric $V$ are also solutions of \eqref{eq:vergeo}.  
Although it looks nonlinear, \eqref{eq:vergeo} actually has a simple 
structure.  Let $W$ and $C$ be the symmetric and skew part of $V$, 
respectively.  As noted above, $C$ must be constant, so 
\eqref{eq:vergeo} becomes
\begin{equation}
	\dot{W}^{ii'} = -2 \sum_{k}\bigl( C^{ik}W^{ki'} + C^{i'k}W^{ki} \bigr),
\end{equation}
i.e., a system of linear ordinary differential equations in $W$.

\subsection{Horizontal geodesics}
\label{sec:horgeo}

If $\gamma$ is horizontal, $V^{ii'}\equiv0$ along $\gamma$.  It 
follows from \eqref{eq:LMgeo2} that $v^i\equiv0$, so there are no 
horizontal geodesics of $(LM,G)$.  In particular, no geodesic of 
$(LM,G)$ is a horizontal lift of a geodesic of $(M,g)$.

In \cite{Stahl:imprisoned}, properties of curves $\gamma$ in $M$ with 
extremal length was studied, with the length measured in a parallel 
pseudo-orthonormal frame along $\gamma$.  It was found that for an 
extremal $\gamma$ to be a geodesic of $(M,g)$, it is necessary that
\begin{equation}
	\delta_{ij} W^i R^j\sb{klm} W^k W^l = 0,
\end{equation}
where $W^i$ is the tangent vector of $\gamma$ and $R^j\sb{klm}$ are 
the components of the Riemann tensor, both expressed in the parallel 
frame.  This is a severe restriction on $(M,g)$.  

Locally, geodesics may be considered as extremal curves of the length 
functional.  We can reformulate the result in \cite{Stahl:imprisoned} 
as follows: even if we restrict attention to horizontal curves 
in $OM$, an extremal of the \b{length} functional cannot be expected to 
be a lift of a geodesic in $(M,g)$.

\section{Discussion}
\label{sec:discussion}

There are many open questions as to the structure of $(LM,G)$ for 
physically interesting spacetimes $(M,g)$.  In fact, even for the 
2-dimensional Schwarzschild spacetime, it is not known if the 
\b{boundary} is of dimension 0 or 1.  Hopefully, geometric methods in 
$(LM,G)$ may help to shed some light on the situation.  But it should 
be remembered that actual calculations with the \b{boundary} is 
difficult even in the two-dimensional case.

Proposition~\ref{pr:scalardiv} also raises some questions.  Loosely 
speaking, that $\tilde{R}\to-\infty$ may be interpreted as that the 
geometry becomes `infinitely hyperbolic' at the boundary point.  
However, not all sectional curvatures have to diverge, which means 
that the `circumference' of the boundary point may be finite in some 
planes while it is infinite in other planes.

Also, the geodesics of $(LM,G)$ may be very complicated in general, 
with no obvious connection to the geodesics of $(M,g)$.  The geodesic 
equations (\ref{eq:LMgeo1}--\ref{eq:LMgeo2}) suggest the possibility 
of a kind of oscillatory behaviour of the horizontal components 
$v^i$ of the tangent vector.

\section*{Acknowledgements}
The calculations of the connection and curvature forms were verified 
with a modified version of the \emph{Ricci} package \cite{Lee:Ricci} 
for \emph{Mathematica} (Wolfram Research, Inc).

\appendix
\section{Expressions for the curvature}
\label{app:curvature}

As described in \S\ref{sec:connection}, the curvature form 
$\curvformLM$ of $(LM,G)$ can be found by applying 
(\ref{eq:Cartan1M}--\ref{eq:Cartan2M}) and Lemma~\ref{la:dR} to 
\eqref{eq:connLMsoln}.  Because of the different metrics involved, 
some care is needed to keep track of which metric is used for 
contractions and raising and lowering of indices.  Note that covariant 
and contravariant components in the frame 
$E_I=(\canform^i,\connform^i\sb{i'})$ may be identified, since the 
frame components of $G$ is given by
\begin{equation}
	G_{ij}=\delta_{ij},\quad
	G_{i\{j,j'\}}=G_{\{j,j'\}i}=0\quad\text{and}\quad
	G_{\{i,i'\}\{j,j'\}}=\delta_{ij}\delta_{i'j'}.
\end{equation}
So we may use $\delta$, the Kronecker delta, for index operations.  We 
also use the notation $\canform^i$, $\connform^i\sb{j}$ and 
$R^i\sb{jkl}$ for the components of the canonical form $\canform$, the 
connection form $\connform$ and the curvature tensor $R$ of $(M,g)$ in 
the frame on $(M,g)$ specified by the location in $LM$.  If we raise 
and lower indices with $\delta$, an ambiguity arises when applying 
symmetries and contractions to the Riemann tensor.  For example, 
$R_{ijkl}$ will \emph{not} be equal to $R_{jikl}$ since the first 
index is lowered with $\delta$ instead of $g$.  Therefore we keep the 
index positions fixed and adopt the convention that all repeated 
indices represent contractions, regardless of their variance.  We 
content ourselves with showing the results, since the calculations 
involve a substantial amount of algebra.

The components of the curvature form $\curvformLM$ in the frame 
$E_I=(\canform^i,\connform^i\sb{i'})$ are
\begin{equation}\label{eq:Omega1}
\begin{split}
\curvformLM^i\sb{j} &= 
		\tfrac12\, \canform^i \!\wedge \canform^j \\
	&  + \tfrac14\, \bigl(
				R^i\sb{lkj} - R^j\sb{lki} + R^k\sb{lij} 
			- R^{mm'}\sb{ik} R_{mm'jl} - R^{mm'}\sb{ij} R_{mm'kl} 
			\bigr)\, \canform^k \!\wedge \canform^l \\
  & - \tfrac12\, R^l\sb{l'ij;k}\, \canform^k \!\wedge \connform^l\sb{l'}
    - \tfrac14\, 
				\bigl( \connform^i\sb{k} + \connform^k\sb{i} \bigr)
				\wedge
				\bigl( \connform^j\sb{k} + \connform^k\sb{j} \bigr) \\
  & + \tfrac12\, R^k\sb{lij} \bigl(
				\connform^m\sb{k} \wedge \connform^m\sb{l}
			+ \connform^k\sb{m} \wedge \connform^m\sb{l}
			+ \connform^k\sb{m} \wedge \connform^l\sb{m}
			\bigr) \\
	& + \tfrac14\, R^k\sb{lim} \connform^k\sb{l} \wedge 
				\bigl( \connform^m\sb{j} + \connform^j\sb{m} \bigr) 
	  + \tfrac14\, R^k\sb{ljm} \connform^k\sb{l} \wedge
				\bigl( \connform^m\sb{i} + \connform^i\sb{m} \bigr) \\
  & - \tfrac14\, R^k\sb{k'im} R^l\sb{l'jm}\, 
				\connform^k\sb{k'} \wedge \connform^l\sb{l'},
\end{split}
\end{equation}
\begin{equation}\label{eq:Omega2}
\begin{split}
\curvformLM^i\sb{\{j,j'\}} &= -\curvformLM^{\{j,j'\}}\sb{i} =
		\tfrac12\, R^j\sb{j'ik;l}\, \canform^k \!\wedge \canform^l  \\
	&	+ \tfrac14\, \bigl(
				\canform^{j'} \!\wedge \connform^j\sb{i}
			- \canform^j \!\wedge \connform^i\sb{j'}
			- \delta_{ij}\, \canform^k \!\wedge 
					( \connform^{j'}\sb{k} + \connform^k\sb{j'} )
			- \delta_{ij'}\, \canform^k \!\wedge \connform^k\sb{j}
			\bigr) \\
	&	- \tfrac14\, \bigl(
				R^l\sb{l'ij'}\, \canform^j 
			+ R^l\sb{l'ij}\, \canform^{j'}
			\bigr) \wedge \connform^l\sb{l'}
		- \tfrac14\, R^j\sb{j'kl}\, \canform^k \!\wedge 
				\bigl( \connform^i\sb{l} + \connform^l\sb{i} \bigr) \\
	& - \tfrac14\, R^l\sb{j'ik} \canform^k \!\wedge
				\bigl( \connform^j\sb{l} + \connform^l\sb{j} \bigr) 
		+ \tfrac14\, R^j\sb{lik}\, \canform^k \!\wedge 
				\bigl( \connform^{j'}\sb{l} + \connform^l\sb{j'} \bigr) \\
	&	+ \tfrac14\, R^{j'}\sb{lik}\, \canform^k \!\wedge \connform^j\sb{l} 
		- \tfrac14\, R^l\sb{jik}\, \canform^k \!\wedge \connform^l\sb{j'} 
		- \tfrac14\, R^l\sb{l'im} R^j\sb{j'mk}\, 
				\canform^k \!\wedge \connform^l\sb{l'},
\end{split}
\end{equation}
\begin{equation}\label{eq:Omega3}
\begin{split}
\curvformLM&^{\{i,i'\}}\sb{\{j,j'\}} =
		- \tfrac14 \bigl(
				\delta_{i'j'}\, \canform^i \!\wedge \canform^j
			+	\delta_{i'j}\, \canform^i \!\wedge \canform^{j'}
			+	\delta_{ij'}\, \canform^{i'} \!\wedge \canform^j
			+	\delta_{ij}\, \canform^{i'} \!\wedge \canform^{j'}
			\bigr) \\
	&	+ \tfrac14\, \bigl(
				R^i\sb{i'j'l}\, \canform^j + R^i\sb{i'jl}\, \canform^{j'}
			- R^j\sb{j'i'l}\, \canform^i - R^j\sb{j'il}\, \canform^{i'}
			\bigr) \wedge \canform^l \\
	&	+ \tfrac14\, \bigl(
				\delta_{i'j'} ( R^i\sb{jkl} -  R^j\sb{ikl} )
			+ \delta_{ij} ( R^{i'}\sb{j'kl} - R^{j'}\sb{i'kl} )
			+ \delta_{i'j} R^i\sb{j'kl} - \delta_{j'i} R^j\sb{i'kl}
			\bigr)\, \canform^k \!\wedge \canform^{l} \\
	&	- \tfrac14\, R^i\sb{i'km} R^j\sb{j'lm}\, \canform^k \!\wedge \canform^l 
		+ \tfrac14\, \bigl(
				\connform^i\sb{j} \wedge \connform^{i'}\sb{j'}
			- \connform^j\sb{i} \wedge \connform^{j'}\sb{i'}
			\bigr) \\
	&	- \tfrac14\, \delta_{ij}\, \bigl(
				( \connform^{i'}\sb{k} + \connform^k\sb{i'} ) \wedge
				( \connform^k\sb{j'} + \connform^{j'}\sb{k} )
			+ \connform^k\sb{i'} \wedge \connform^k\sb{j'}
			\bigr) \\
	&	- \tfrac14\, \delta_{i'j'}\, \bigl(
				( \connform^i\sb{k} + \connform^k\sb{i} ) \wedge
				( \connform^k\sb{j} + \connform^j\sb{k} )
			+ \connform^i\sb{k} \wedge \connform^j\sb{k}
			\bigr) \\
	&	- \tfrac14\, \delta_{ij'}\, \bigl(
				\connform^{i'}\sb{k} \wedge \connform^j\sb{k}
			+ \connform^k\sb{i'} \wedge \connform^k\sb{j}
			\bigr) 
		- \tfrac14\, \delta_{i'j}\, \bigl(
				\connform^i\sb{k} \wedge \connform^{j'}\sb{k}
			+ \connform^k\sb{i} \wedge \connform^k\sb{j'}
			\bigr).
\end{split}\raisetag{23mm}
\end{equation}

From (\ref{eq:Omega1}--\ref{eq:Omega3}) we can obtain the components 
of the Riemann tensor, and a contraction then gives the Ricci tensor:
\begin{align}
	\label{eq:Ricci1}
	\tilde{R}_{ij} &= 
		-\,\delta_{ij} - \tfrac12\, R^k\sb{k'il} R^k\sb{k'jl} + R_{ij}, \\
	\label{eq:Ricci2}
	\tilde{R}_{i\{j,j'\}} &= \tilde{R}_{\{j,j'\}i} =
		\tfrac12\, R^j\sb{j'ik;k}, \\
	\label{eq:Ricci3}
	\tilde{R}_{\{i,i'\}\{j,j'\}} &= 
	-\, \tfrac{2n+1}2\, \delta_{ij'} \delta_{i'j} 
	- \tfrac{n+1}2\, \delta_{ij} \delta_{i'j'}
	+ \tfrac32\, \delta_{ii'}\delta_{jj'}
	+ \tfrac14\, R^i\sb{i'kk'} R^j\sb{j'kk'}.
\end{align}
Contracting again gives the curvature scalar $\tilde{R}$ as given in 
\S\ref{sec:curvature}, equation \eqref{eq:curvscalarLM}.

\providecommand{\bysame}{\leavevmode\hbox to3em{\hrulefill}\thinspace}

\end{document}